\documentclass[conference,10pt]{IEEEtran}
\usepackage{amsmath}
\usepackage{amssymb}
\usepackage{graphicx}
\usepackage{color}
\usepackage{cite}
\usepackage{bm}
\usepackage{epsfig,psfrag}
\usepackage{tabularx}
\usepackage{acronym}
\usepackage[yyyymmdd,hhmmss]{datetime}
\usepackage{bbm}
\usepackage{mathrsfs} 
\usepackage{tabularx}
\usepackage{multirow}
\usepackage{colortbl,pgfplotstable}
\usepackage{subfig}
\usepackage{tabularx}
\usepackage{multirow}
\usepackage{acronym}

\acrodef{spa}[SPA]{sum-product algorithm}
\acrodef{da}[DA]{data association}
\acrodef{mmse}[MMSE]{minimum mean-square error}
\acrodef{pt}[PT]{potential target}
\acrodef{pmf}[pmf]{probability mass function}
\acrodef{pdf}[pdf]{probability density function}
\acrodef{iid}[iid]{independent and identically distributed}
\acrodef{rmse}[RMSE]{root-mean-squared error}
\acrodef{ospa}[OSPA]{optimal sub-pattern assignment}
\acrodef{bp}[BP]{belief propagation}
\acrodef{bpf}[BPF]{bootstrap particle filter}
\acrodef{upf}[UPF]{unscented particle filter}
\acrodef{pde}[PDE]{partial differential equation}
\acrodef{sde}[SDE]{stochastic differential equation}
\acrodef{ode}[ODE]{ordinary differential equation}
\acrodef{edh}[EDH]{exact Daum and Huang}
\acrodef{ledh}[LEDH]{localized exact Duam and Huang}
\acrodef{pfpf}[PFPF]{particle flow particle filter}
\acrodef{mcmc}[MCMC]{Markov Chain Monte Carlo}
\acrodef{smc}[SMC]{sequential Monte Carlo}
\acrodef{map}[MAP]{maximum a posteriori}
\acrodef{tdoa}[TDOA]{Time-difference of arrival}
\acrodef{pf}[PF]{Particle flow}
\acrodef{pda}[PDA]{probabilistic data association}
\acrodef{jpda}[JPDA]{Joint \ac{pda}}
\acrodef{phd}[PHD]{probability hypothesis density}
\acrodef{cphd}[CPHD]{cardinalized \ac{phd}}
\acrodef{mht}[MHT]{multi-hypothesis tracking}
\acrodef{slam}[SLAM]{simultaneous localization and mapping}
\acrodef{iid}[iid]{independent and identically distributed}
\acrodef{rfs}[RFS]{random finite sets}
\acrodef{gospa}[GOSPA]{general optimal subpattern assignment}

\newcommand{\rv}[1]{\mathsf{#1}}  
\newcommand{\V}[1]{\bm{#1}}
\newcommand{\RV}[1]{\boldsymbol{\mathsf{#1}}}

\newcommand{\M}[1]{\boldsymbol{\uppercase{#1}}}

\newcommand{\Set}[1]{\mathcal{#1}}

\newcolumntype{L}[1]{>{\raggedright\arraybackslash}p{#1}}
\newcolumntype{C}[1]{>{\centering\arraybackslash}p{#1}}
\newcolumntype{R}[1]{>{\raggedleft\arraybackslash}p{#1}}

\providecommand{\be}{\begin{equation}}
\providecommand{\ee}{\end{equation}}

\providecommand{\ist}{\hspace*{.3mm}}

\providecommand{\rmv}{\hspace*{-.3mm}}

\providecommand{\nn}{\nonumber}
\newcommand{\T}{\text{T}}
\DeclareMathOperator*{\argmax}{arg\,max}

\allowdisplaybreaks
\definecolor{temporalgreen}{RGB}{0,128,0}
\definecolor{spatialred}{RGB}{255,0,0}
\definecolor{temporalblue}{RGB}{0,0,205}

\allowdisplaybreaks

\begin{document}
\title{Track Coalescence and Repulsion:\\ MHT, JPDA, and BP}

\author{\IEEEauthorblockN{Thomas Kropfreiter\IEEEauthorrefmark{1}, Florian Meyer\IEEEauthorrefmark{1}, Stefano Coraluppi\IEEEauthorrefmark{2}, Craig Carthel\IEEEauthorrefmark{2}, Rico Mendrzik\IEEEauthorrefmark{3},  and Peter Willett\IEEEauthorrefmark{4} \vspace{2mm}}
\IEEEauthorblockA{\IEEEauthorrefmark{1}Scripps Institution of Oceanography and Department of Electrical and Computer Engineering} \\[-4.6mm]
\IEEEauthorblockA{University of California San Diego, La Jolla, CA, USA (\{flmeyer, tkropfreiter\}@ucsd.edu) \vspace{1mm}}

\IEEEauthorblockA{\IEEEauthorrefmark{2}Systems \& Technology Research, Woburn, MA, USA (\{stefano.coraluppi, craig.carthel\}@stresearch.com) \vspace{1mm}} 

\IEEEauthorblockA{\IEEEauthorrefmark{3}Ibeo Automotive Systems GmbH, Hamburg, Germany (rico.mendrzik@ibeo-as.com) \vspace{1mm}} 

\IEEEauthorblockA{\IEEEauthorrefmark{4}Department of ECE, University of Connecticut, Storrs, CT, USA (peter.willett@uconn.edu)}  \vspace{-6.5mm}}

\maketitle

\begin{abstract}
Joint probabilistic data association (JPDA) and multiple hypothesis tracking (MHT) introduced in the 70s, are still widely used methods for multitarget tracking (MTT). Extensive studies over the last few decades have revealed undesirable behavior of JPDA and MHT type methods in tracking scenarios with targets in close proximity.  In particular, JPDA suffers from the track coalescence effect, i.e., estimated tracks of targets in close proximity tend to merge and can become indistinguishable. Interestingly, in MHT, an opposite effect to track coalescence called track repulsion can be observed. In this paper, we review the estimation strategies of the MHT, JPDA, and the recently introduced belief propagation (BP) framework for MTT and we investigate if BP also suffers from these two detrimental effects. Our numerical results indicate that BP-based MTT can mostly avoid both track repulsion and coalescence.

\vspace{3mm}\end{abstract}


\acresetall
\section{Introduction}\label{sec:intro}

Multitarget tracking (MTT) \cite{barShalom11,ChaMor11,mahler2007statistical,koch14,mey18} is a key signal processing task that enables situational awareness in a variety of applications. Historically, the most important MTT scenarios are aerospace surveillance \cite{reid79,barShalom74,bar-shalom09} and applied ocean sciences \cite{fortmann83,FerMunTesBraMeyPelPetAlvStrLeP:J17,Kim21}. However, in the context of autonomous navigation, robotic applications are recently drawing increased attention \cite{levinson11,MeyWil:J21}. MTT methods are typically derived in the Bayesian estimation framework and they rely on different modeling paradigms, which can be broadly classified as vector-type \cite{musicki04,VerMas05,MusLaS08,horridge06,PatPop00,Cox96,DanNew06,Kur90,Bla04,Mor77,deb97} and set-type  \cite{Mah03,vo05,vo07,williams2015marg,Nannuru16}. 

Vector-type MTT methods describe both multitarget states and measurements by random vectors.  The well-known joint probabilistic data association (JPDA) filter \cite{barShalom74} introduced in the 70s established the probabilistic \ac{da} paradigm for MTT. The JPDA aims to calculate the minimum mean square error (MMSE) state estimates  for a known number of targets. Data association hypotheses are employed in a soft manner by performing weighted sums over all possible associations at each time step. Performing these weighted sums is equivalent to calculate the marginal posterior probability density functions needed for state estimation by ``marginalizing out'' the unknown data association vector. In contrast, the very popular multiple hypothesis tracking (MHT) methods \cite{reid79} seek to calculate the maximum a posteriori (MAP) estimate of an entire sequence of data association vectors. In particular, MHT methods perform hard data association based on the single most likely measurement-target combination over multiple consecutive instances of time \cite{reid79}. After an approximate MAP estimate of the data association vectors has been fixed, MAP or MMSE estimates of the target states can be calculated individually by means of sequential Bayesian estimation, e.g., extended Kalman filtering.

Extensive studies over the last few decades have revealed undesirable behavior of JPDA and MHT type methods in tracking scenarios with targets in close proximity. In particular, JPDA type methods suffer from the track coalescence effect \cite{fitzgerald85}, i.e., estimated tracks of closely spaced targets that move in parallel will tend to come together, merge, and can become indistinguishable. Interestingly, in  MHT, an opposite effect to track coalescence called track repulsion \cite{corcar09}  can be observed. In particular, estimated tracks of close targets that move in parallel repel each other in the sense that their separation is larger compared to the distance between the actual targets. Methodological adaptations tailored to alleviate these detrimental effects have been introduced in  \cite{bloblo00,bloblo15,corcar12,SveSve11}. Track coalescence can be avoided by using a special hypothesis pruning strategy \cite{bloblo00,bloblo15} and track repulsion by selecting a global hypothesis that is not the single MAP solution, but an element of a MAP equivalence class  \cite{corcar12}. While these adaptations can reduce the effects of track coalescence and repulsion, they are unsuitable for large-scale tracking scenarios with an unknown and time-varying number of targets.

Set-type methods describe both multitarget states and measurements by random finite sets. Some approaches in this class including probability hypothesis density (PHD) methods  \cite{Mah03,vo05,vo07,Nannuru16} avoid data association at the cost of a reduced target detection and tracking performance as well as a suboptimal postprocessing step for track formation. Some other RFS methods rely on association strategies similar to the ones performed by the JPDA filter and thus also suffer from track coalescence \cite{reuter14,williams2015marg}.

 Recently developed graph signal processing methods for MTT \cite{williams14,mey17,mey18,MeyWil:J21} fit in the vector-type family. Here, filtering and data association for randomly appearing and disappearing targets are described by a joint factor graph. This factor graph provides the blueprint for sum-product message passing algorithms also known as belief propagation (BP). By following the MMSE estimation approach of the JPDA, BP performs efficient marginalization operations and can provide scalable solutions to two pertinent problems: the data association problem that arises due to measurement-origin uncertainty and the presence of clutter as well as the nonlinear filtering problem with appearing and disappearing target states.
 
 In this paper, we review the track coalescence and repulsion effects of the JPDA and MHT methods and investigate them in the context of BP-based MTT. The contributions of this paper are as \vspace{.1mm} follows.
\begin{itemize}
\item We review the estimation strategies of the MHT, JPDA, and BP approaches to MTT.
\vspace{.5mm}
\item We discuss the track coalescence and track repulsion effects of MHT, JPDA, and BP.
\vspace{.5mm}
\item We demonstrate significantly reduced track coalescence and repulsion of BP in three different case studies.
\end{itemize}

\emph{Notation:} Random variables are displayed in sans serif, upright fonts; their realizations in serif, italic fonts. 
Vectors and matrices are denoted by bold lowercase and uppercase letters, respectively. For example, a random variable and its realization are denoted by $\rv x$ and $x$ and a random vector and its realization 
by $\RV x$ and $\V x$.
Furthermore, ${\V{x}}^{\text T}$ denotes the transpose of vector $\V x$; 
$\propto$ indicates equality up to a normalization factor;
$f(\V x)$ denotes the \ac{pdf} of 
random  vector $\RV{x}$. Finally, $1(a)$ denotes the unit sample function (i.e., $1(a) \rmv=\rmv 1$ if $a \rmv=\rmv 0$ and $0$\vspace{-.5mm} otherwise).

\section{General System Model}\label{sec:systemModel}
\vspace{-.5mm}

In this section, we introduce a general MTT system model.
\vspace{-5mm}

\subsection{Potential Target States and State-Transition PDF} 
In most practical applications, the number of targets is time-varying and unknown. Here, we introduce \ac{pt} states \cite{mey17,mey18}. At discrete time $k \rmv\geq\rmv 0$, the number of \acp{pt} $j_k$ is the maximum possible number of targets that have generated a measurement. The state of \ac{pt} $j \in \{1,\dots,j_k\}$ consists of a binary existence variable $\rv{r}^{(j)}_{k} \rmv\in\rmv \{0,1\}$ and a kinematic state $\RV{x}^{(j)}_{k}\rmv\in\rmv \mathbb{R}^{n}\rmv$. The existence variable $\rv{r}^{(j)}_{k}$ models the existence/nonexistence of \ac{pt} $j$ in the sense that \ac{pt} $j$ exists at time $k$ if and only if $\rv{r}^{(j)}_{k} \!=\! 1$.
The state $\RV{x}^{(j)}_{k}$ of \ac{pt} $j$ consists of the \ac{pt}'s position and possibly further motion-related parameters. 

The kinematic state $\RV{x}^{(j)}_{k}$ of \acp{pt} with $\rv{r}^{(j)}_{k} \!=\! 0$ is not defined and can be represented by an arbitrary ``dummy'' \ac{pdf} $f_{\text{D}}\big(\V{x}^{(j)}_{k}\big)$ with $\int \rmv f_{\text{D}}\big(\V{x}^{(j)}_{k}\big) \mathrm{d} \V{x}^{(j)}_{k} \rmv=\rmv 1$. All \acp{pdf} of \ac{pt} states, $ f\big(\V{x}^{(j)}_{k}\rmv, r^{(j)}_{k}\big)$, have the property that $f\big(\V{x}^{(j)}_{k}\rmv, 0 \big) \rmv=\rmv f^{(j)}_{k} f_{\text{D}}\big(\V{x}^{(j)}_{k}\big)$,  where $f^{(j)}_{k} \!\rmv\rmv\in\rmv\! [0,1]$ is the probability of nonexistence. For each \ac{pt} state $\big(\RV{x}^{(j)}_{k-1}, \; \rv{r}^{(j)}_{k-1} \big)$, $j \in \{1,\dots,j_{k-1}\}$ at time $k\rmv-\rmv1$, there is one ``legacy'' \ac{pt} state $\big( \underline{\RV{x}}^{(j)}_{k}\rmv\rmv, \; \underline{\rv{r}}^{(j)}_{k} \big)$ at time $k$. The motion and disappearance of each \ac{pt} is modeled by the single-target state-transition \ac{pdf}  $f\big( \underline{\V{x}}_k^{(j)}\rmv\rmv, \underline{r}_k^{(j)} \big| \V{x}_{k-1}^{(j)}, r_{k-1}^{(j)}   \big)$ that involves the kinematic state-transition \ac{pdf} $f\big( \underline{\V{x}}_k^{(j)} \big| \V{x}_{k-1}^{(j)} \big)$ and the probability of target survival $p_{\text{s}}$  (e.g., see \cite[Section VIII-C]{mey18}). It is assumed that each target evolves independently.  At time $k \rmv=\rmv 0$, the target states are assumed statistically independent across \acp{pt} $j$. However, often no prior information for \ac{pt} states is available, i.e., $j_0 \rmv=\rmv 0$.
\vspace{-1mm}

\subsection{New \acp{pt}, Data Association, and Measurement Likelihood Function}\label{sec:DA-MLF}
\vspace{-.5mm}

At each time (or ``scan'') $k\rmv\geq\rmv1$, a sensor produces measurements $\RV{z}^{(m)}_{k} \rmv$, $m \rmv\in\rmv \big\{ 1,\dots,\rv{m}_k \big\}$. (Note that up to time $k$, the number of measurements $\rv{m}_k$ is random.) Each measurement can have one of three possible sources: (i) a new \ac{pt} representing a target that generates a measurement for the first time;  (ii) a legacy \ac{pt} representing a target that has generated at least one measurement before; and (iii) clutter. At time $k$, $\rv{m}_k$ new \ac{pt} states are introduced, i.e., $\big(\overline{\RV{x}}^{(m)}_{k} \; \overline{\rv{r}}^{(m)}_{k} \big)$, $m \in \{1,\dots,\rv{m}_k\}$. Here, $\overline{\rv{r}}^{(m)}_{k} = 1$ means that measurement $\RV{z}^{(m)}_{k}$ has source (i) and $\overline{\rv{r}}^{(m)}_{k} = 0$ means that measurement $\RV{z}^{(m)}_{k}$ has either source (ii) or (iii). The birth of new targets is modeled by a Poisson point process with mean $\mu_{\text{b}}$ and \ac{pdf} $f_{\text{b}}\big( \overline{\V{x}}_{k} \big)$.  The joint vectors of all existence variables and kinematic states at time $k$ are denoted as $\RV{x}_k = \big[\underline{\RV{x}}^{(1)}_k \cdots \underline{\RV{x}}^{(j_{k-1})}_k \ist\ist \overline{\RV{x}}^{(1)}_k \cdots \overline{\RV{x}}^{(m_k)}_k\big]^{\mathrm{T}}$ and $\rv{r}_k = \big[\underline{\rv{r}}^{(1)}_k \cdots \underline{\rv{r}}^{(j_{k-1})}_k \ist\ist \overline{\rv{r}}^{(1)}_k \cdots \overline{\rv{r}}^{(m_k)}_k\big]^{\mathrm{T}}\rmv\rmv\rmv$, respectively.

The target represented by a \ac{pt} $j$ is detected (in the sense that it generates a measurement $\RV{z}_k^{(m)}$) with probability $p_{\text{d}}$. The statistical relationship of a measurement $\RV{z}_k^{(m)}$ and the kinematic state $\RV{x}_k^{(j)}$  of the corresponding detected \ac{pt}, is described by the conditional \ac{pdf} $f\big(\V{z}^{(m)}_{k} | \V{x}^{(j)}_{k} \big)$. This conditional \ac{pdf} can be derived from the measurement model of the sensor. Clutter measurements are modeled by a Poisson point process with mean $\mu_{\text{c}}$ and \ac{pdf} $f_{\text{c}}\big( \V{z}^{(m)}_{k} \big)$. At time $k$, the joint measurement vector is fixed and denoted as $\V{z}_k \!\triangleq\rmv \big[\V{z}_{k}^{(1)\ist\T} \rmv\cdots\ist \V{z}_{k}^{(m_k)\ist\T}\big]^{\T}\rmv\rmv.$ Thus, the total number of (legacy and new) \acp{pt} is $j_{k} \!=\rmv j_{k-1} \!+\rmv m_{k}$.

In MTT, the source of measurements is uncertain. In particular, it is unknown which measurement is generated by which \ac{pt}, and a measurement may also be clutter, i.e., not generated by any \ac{pt}. The considered MTT problem follows the conventional \ac{da} assumption, i.e., a target can originate at most one measurement and a measurement can be originated from at most one target \cite{barShalom11,mahler2007statistical,mey18}.
The association between $m_k$ measurements and $j_{k-1}$ legacy \acp{pt} at time $k$ is typically modeled by a ``target-oriented'' \ac{da} vector $\RV{a}_{k} = \big[\rv{a}_{k}^{(1)} \cdots\ist \rv{a}_{k}^{(j_{k-1})} \big]^{\T}\rmv\rmv$. The target-oriented association variable $\rv{a}_{k}^{(j)}$ is $m \in \{1,\dots,m_{k} \}$ if \ac{pt} $j$ originated measurement $m$ and zero if \ac{pt} $j$ did not originate any measurement \cite{barShalom11,mey18}. A DA vector $\V{a}_{k}$ is valid if it does not violate the conventional DA association assumption.

We also introduce the ``measurement-oriented'' \ac{da} vector $\RV{b}_{k} = \big[\rv{b}_{k}^{(1)} \cdots\ist \rv{b}_{k}^{(m_k)} \big]^{\T}\rmv\rmv$. The measurement-oriented association variable $\rv{b}_{k}^{(m)}$ is $j \in \{1,\dots,j_{k-1} \}$ if measurement $m$ is generated by legacy \ac{pt} $j$ and zero if it is generated by clutter or by a newly detected \vspace{1.5mm}\ac{pt}. The\vspace{-1.5mm} DA representation by means of both target-oriented and measurement-oriented DA vectors is redundant in the sense that for any valid $\V{a}_{k}$, there is exactly one valid $\V{b}_{k}$ and for any valid $\V{b}_{k}$, there is exactly one valid $\V{a}_{k}$. However, this representation based on both $\V{a}_{k}$ and $\V{b}_{k}$ makes it possible to develop scalable MTT methods that exploit the structure of the DA problem to reduce computational complexity.
\vspace{-1mm}

\subsection{Joint Posterior PDF and Factor Graph}
\label{eq:JointAssociation}

Following common assumptions \cite{barShalom11,reid79,mahler2007statistical,williams2015marg,williams14,meyer16,mey17,mey18}, the joint posterior \ac{pdf} of $\RV{x}_{0:k}$, $\RV{r}_{0:k}$, $\RV{a}_{1:k}$, and $\RV{b}_{1:k}$ conditioned on fixed measurements $\V{z}_{1:k}$ can be obtained \vspace{.5mm}as
\begin{align}
  &f\big( \V{x}_{0:k}, \V{r}_{0:k}, \V{a}_{1:k}, \V{b}_{1:k} \big| \V{z}_{1:k} \big) \nn \\[0.5mm]
  &\hspace{0mm} \propto  \Bigg(\prod^{j_{0}}_{j''=1} f\Big(\V{x}^{(j'')}_{0}, r^{(j'')}_{0}  \Big)  \Bigg) \prod^{k}_{k'=1}  \nn \\[0.5mm]
  &\hspace{4.5mm}\times\hspace{1mm} \Bigg(\prod^{j_{k'-1}}_{j'=1} f\big(\underline{\V{x}}^{(j')}_{k'},\underline{r}^{(j')}_{k'}\big|\V{x}^{(j')}_{k'-1},r^{(j')}_{k'-1}\big)  \Bigg)   \nn\\[0.5mm] 
  &\hspace{4.5mm}\times \Bigg( \prod^{j_{k'-1}}_{j=1}  q\big( \underline{\V{x}}^{(j)}_{k'}\!, \underline{r}^{(j)}_{k'}\!\rmv, a^{(j)}_{k'}\rmv; \V{z}_{k'} \big)\rmv\rmv \prod^{m_{k'}}_{m'=1} \Psi_{j\rmv,m'}\big(a_{k'}^{(j)}\rmv,b_{k'}^{(m')}\big) \rmv\Bigg)  \nn\\[1.5mm]
  &\hspace{4.5mm}\times  \hspace{1mm} \prod^{m_{k'}}_{m=1} \rmv v\big( \overline{\V{x}}^{(m)}_{k'}\!, \overline{r}^{(m)}_{k'}\!, b^{(m)}_{k'}\rmv; \V{z}_{k'}^{(m)} \big)\hspace{-.1mm}.
  \label{eq:jointPosteriorComplete} \\[-6mm]
  \nn
\end{align}
Here, the new \ac{pt} factor $v\big( \overline{\V{x}}^{(m)}_{k}\!, \overline{r}^{(m)}_{k},$ $b^{(m)}_{k}\rmv; \V{z}_{k}^{(m)} \big)$ is given\vspace{.5mm} by
\begin{align}
&\hspace{-.9mm}v\big( \overline{\V{x}}^{(m)}_{k}\!, 1, b^{(m)}_{k}\rmv; \V{z}_{k}^{(m)} \big) \nn \\[.5mm]
&\hspace{-.9mm}\triangleq \begin{cases}
     0 \ist,  & \hspace{-1mm} b^{(m)}_{k} \rmv\in\rmv \{ 1,\dots,j_{k-1} \}\\[2mm]
     { \ist \frac{p_{\mathrm{d}} \ist \mu_{\text{b}} \ist f_{\text{b}}\big(\overline{\V{x}}^{(m)}_{k}\big)}{ \mu_{\text{c}} f_{\text{c}}\big( \V{z}_{k}^{(m)} \big)} }  f\big(\V{z}_k^{(m)}  \big| \overline{\V{x}}^{(m)}_{k} \big) \ist, & \hspace{-1mm} b^{(m)}_{k} \!=\rmv 0 
  \end{cases}
  \nn\\[-3.5mm]
  \nn
\end{align}
and $v\big( \overline{\V{x}}^{(m)}_{k}\!, 0, b^{(m)}_{k}\rmv; \V{z}_{k}^{(m)} \big) \rmv\triangleq\rmv f_{\text{D}}\big(\overline{\V{x}}^{(m)}_{k}\big)$. In addition, the legacy \ac{pt} factor $q\big( \underline{\V{x}}^{(j)}_{k}\!, \underline{r}^{(j)}_{k},$ $a^{(j)}_{k};\V{z}_{k} \big)$ \vspace{.5mm} reads
\begin{align}
&\hspace{-.8mm}q\big( \underline{\V{x}}^{(j)}_{k}\!, 1, a^{(j)}_{k}\rmv; \V{z}_{k} \big) \nn\\
&\hspace{-.8mm}\triangleq \begin{cases}
    \frac{ p_{\text{d}} }{ \mu_{\text{c}} f_{\text{c}}\big( \V{z}_{k}^{(m)} \big)} f\big( \V{z}_{k}^{(m)} \rmv\big|\ist \underline{\V{x}}_{k}^{(j)} \big), & a^{(j)}_{k} \rmv= m \rmv\in\rmv \{1,\dots,m_k \}  \\[2mm]
     1 - p_{\text{d}}  \ist, & a^{(j)}_{k} \rmv=\rmv 0 \ist
  \end{cases}
  \nn
  \\[-5mm]
  \nn
\end{align}
and $q\big( \underline{\V{x}}^{(j)}_{k}\!, 0, a^{(j)}_{k}\rmv; \V{z}_{k} \big) \triangleq 1(a^{(j)}_{k})$. 

Finally, consistency of a pair of target-oriented and measurement-oriented variables $\big(a_{k}^{(j)}\rmv,b_{k}^{(m)}\big)$ is checked by the binary indicator function $\Psi_{j\rmv,m}\big(a_{k}^{(j)}\rmv,b_{k}^{(m)}\big)$. In particular, $\Psi_{j\rmv,m}\big(a_{k}^{(j)}\rmv,b_{k}^{(m)}\big)$ is zero if 
$a_{k}^{(j)} \rmv= m, b^{(m)}_{k} \rmv\neq\rmv j$ or $b^{(m)}_{k} \rmv= j, a_{k}^{(j)} \rmv\neq\rmv m$ and one otherwise (see \cite{williams14,mey18} for details). A detailed derivation of the joint posterior in \eqref{eq:jointPosteriorComplete} has been introduced in \cite[Section~VIII-G]{mey18}.  

After marginalizing out $\RV{b}_{1:k}$ in \eqref{eq:jointPosteriorComplete}, the resulting posterior distribution 
\begin{align}
f\big( \V{x}_{0:k}, \V{r}_{0:k}, \V{a}_{1:k}, \big| \V{z}_{1:k} \big) = \sum_{\V{b}_{1:k}} f\big( \V{x}_{0:k}, \V{r}_{0:k}, \V{a}_{1:k}, \V{b}_{1:k} \big| \V{z}_{1:k} \big) \nn\\[-3.5mm]
 \label{eq:jointPosteriorComplete1} 
\end{align}
represents a system model that is identical to the one used by the MHT\footnote{In the MHT literature, the two vectors $\RV{r}_{0:k}$ and $\RV{a}_{1:k}$ are typically represented by an equivalent single vector $\RV{q}_{0:k}$ referred to as global hypothesis.} and a more general version of the one used by JPDA methods.  Note that this marginalization is trivial since for $\V{a}_{1:k}$ there is exactly one $\V{b}_{1:k}$ for which \eqref{eq:jointPosteriorComplete} is nonzero.

\section{Multitarget Tracking Strategies}\label{sec:Methods}

In this section, we will review the different estimation strategies of the MHT, the JPDA, and the BP methods and discuss how they can cause track repulsion and track coalescence.

\subsection{MHT for Sequence MAP Estimation}
\label{sec:MHT-Tracking}

MHT tracking aims to compute the sequence MAP estimate of  $\RV{x}_{0:k}, \RV{r}_{0:k}, \RV{a}_{1:k}$ given $\V{z}_{1:k}$. In particular, first the sequence MAP of $\RV{r}_{0:k}, \RV{a}_{1:k}$ given $\V{z}_{1:k}$ is computed\vspace{1mm} as 
\begin{align}
&\big(\hat{\V{r}}^{\mathrm{MAP}}_{0:k}\rmv\rmv\rmv,\hat{\V{a}}^{\mathrm{MAP}}_{1:k} \big) = \argmax_{\V{r}_{0:k},\V{a}_{1:k}} \int f\big( \V{x}_{0:k}, \V{r}_{0:k}, \V{a}_{1:k} \big| \V{z}_{1:k} \big) \ist \text{d}\V{x}_{0:k} \nn \\[-2mm]
 \label{eq:MHT0} \\[-6.5mm]
\nn
\end{align}  
This is equivalent to searching for the most likely target-measurement association or global hypothesis \cite{reid79}. The MHT exploits the fact that the marginalization operation $\int \text{d}\V{x}_{0:k}$ can be performed sequentially and, assuming a linear-Gaussian state-transition and measurement model, also in closed form. The optimal MAP solution however is a ``multi-scan solution'' in the sense that the optimal estimate $\big(\hat{\V{r}}^{\mathrm{MAP}}_{0:k}\rmv\rmv\rmv,\hat{\V{a}}^{\mathrm{MAP}}_{1:k} \big)$ at time $k$ can potentially be completely different than the optimal estimate $\big(\hat{\V{r}}^{\mathrm{MAP}}_{0:k-1}\rmv,\hat{\V{a}}^{\mathrm{MAP}}_{1:k-1} \big)$ at time $k-1$. Thus, for optimal MAP estimation it is necessary to keep track of an exponentially growing number of  global hypotheses $\big(\V{r}_{0:k}\rmv,\V{a}_{1:k} \big)$.

Given the most likely target-measurement association $\big(\hat{\V{r}}^{\mathrm{MAP}}_{0:k}\rmv,\hat{\V{a}}^{\mathrm{MAP}}_{1:k} \big)$, the MHT then computes a MAP estimate of $\RV{x}_{0:k}$\vspace{-2mm} as
\begin{align}
\hat{\V{x}}_{0:k}^{\text{MAP}} &= \ist\argmax_{\V{x}_{0:k}} f\big( \V{x}_{0:k} \big| \hat{\V{r}}^{\mathrm{MAP}}_{0:k}\rmv,\hat{\V{a}}^{\mathrm{MAP}}_{1:k}, \V{z}_{1:k}\big) \nn \\[.5mm]
&= \ist\argmax_{\V{x}_{0:k}} f\big( \V{x}_{0:k}, \hat{\V{r}}^{\mathrm{MAP}}_{0:k}\rmv,\hat{\V{a}}^{\mathrm{MAP}}_{1:k} \big| \V{z}_{1:k} \big)\ist
\label{eq:MHT1}
\end{align}
Note that for $\big(\hat{\V{r}}^{\mathrm{MAP}}_{0:k}\rmv,\hat{\V{a}}^{\mathrm{MAP}}_{1:k} \big)$ fixed, the joint posterior \eqref{eq:jointPosteriorComplete1} can be split into independent posterior distributions (one for each detected target according to $\hat{\V{r}}^{\mathrm{MAP}}_{0:k}\rmv$, cf. \eqref{eq:jointPosteriorComplete}). The joint target state estimate $\hat{\V{x}}_{0:k}^{\text{MAP}}$ is then obtained by performing Bayesian smoothing for each detected target in parallel and independent of all other targets, e.g., by means of the Kalman smoother.

In a practical implementation, MHT computes the MAP estimates of the target-measurement association in \eqref{eq:MHT0} and of the target state in \eqref{eq:MHT1} only over a sliding window of $N$ consecutive times steps in order to reduce the computational burden.
A hard decision is then made only on the association at the oldest time step.
MHT can be formulated in form of hypothesis-oriented MHT \cite{reid79} and track-oriented MHT \cite{Kur90,Bla04,Mor77}.
The computational complexity of the original hypothesis-oriented MHT formulation is still problematic due to the high number of hypotheses. 
However, it can be reduced by discarding unlikely hypotheses using, e.g., 
an efficient $m$-best assignment algorithm \cite{Cox96,DanNew06}.

The more efficient track-oriented MHT methods \cite{Kur90,Bla04,Mor77}
represent association hypotheses via a series of tree structures. 
Here, each tree represents the possible association histories of a single target. 
The most likely hypothesis is then found by choosing a leaf node for each single-target tree in such a way that no measurement is used by more than one target. 
Fast hypothesis-search is enabled by means of combinatorial optimization methods \cite{PatDeb92,deb97,PooRij93,PooGad06}. Track-oriented MHT offers a more compact representation of the data association problem than hypothesis-oriented MHT. The two formulations are equivalent under the assumption that target births and clutter follow a Poisson point process \cite{MorCho:C16}.

A limitation of MHT tracking is known as track repulsion, which can arise when targets come in close proximity. It is a direct consequence of performing hard data association by considering only a single global hypothesis $\big(\hat{\V{r}}^{\mathrm{MAP}}_{0:k}\rmv,\hat{\V{a}}^{\mathrm{MAP}}_{1:k} \big)$ for target state estimation according to \eqref{eq:MHT1}. Estimated tracks are displaced at a greater distance than the targets themselves and thus result in an increased estimation error. This displacement can also lead to track swaps (or identity switches) \cite{CorGriDet:C06,Willet07}. In a crossing target scenario the corresponding estimated tracks have the tendency to ``bounce'' more often than they cross. A method for mitigating the detrimental effects of track repulsion has been presented in \cite{corcar12}.
\vspace{0mm}

\begin{figure}[t!]
\centering\hspace{4.5mm}
\vspace{-1mm}
\centering
\includegraphics[scale=.75]{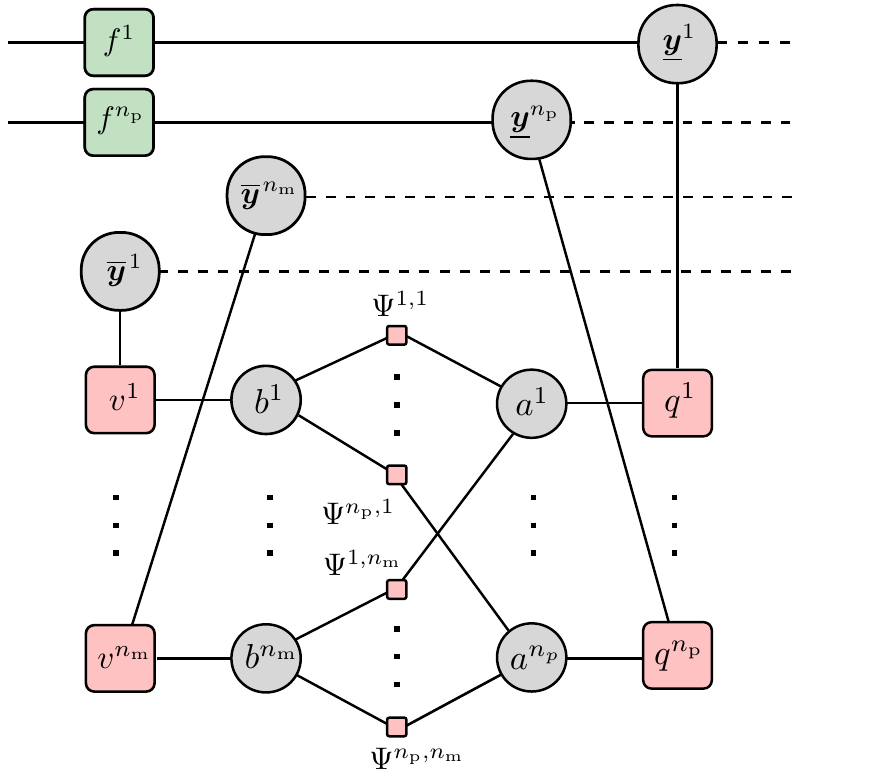}
\vspace{2.5mm}
 \caption{Factor graph for an unknown, time-varying number of objects, corresponding to the joint posterior pdf $f\big( \V{x}_{0:k},\V{r}_{0:k} \V{a}_{1:k}, \V{b}_{1:k} \big| \V{z}_{1:k} \big)$ in \eqref{eq:jointPosteriorComplete}. 
The factor graph depicts one time step $k$. 
The time index $k$ is omitted, and the following short notations are used: 
$n_{\mathrm{m}} \rmv\triangleq m_{k}$, 
$n_{\mathrm{p}} \rmv\triangleq j_{k-1}$,
$\underline{\V{y}}^{j} \rmv\triangleq\rmv [\underline{\V{x}}^{(j)}_k,\underline{r}^{(j)}_k]^{\T}$,
$\overline{\V{y}}^{m} \rmv\triangleq\rmv [\overline{\V{x}}^{(m)}_k,\overline{r}^{(m)}_k]^{\T}$,
$\V{a}^j \rmv\triangleq \V{a}^{(j)}_k$,
$\V{b}^m \rmv\triangleq \V{b}^{(m)}_k$,
$f^j \rmv\triangleq f\big(\underline{\V{x}}^{(j)}_k,\underline{r}^{(j)}_k \big|\V{x}^{(j)}_{k-1},r^{(j)}_{k-1}\big)$,
$q^j \rmv\triangleq q\big( \underline{\V{x}}^{(j)}_{k}\rmv, \underline{r}^{(j)}_{k}\rmv, a_{k}^{(j)};\V{z}_{k} \big)$, 
$v^m \rmv\triangleq v\big( \overline{\V{x}}^{(m)}_{k} \rmv,\overline{r}^{(m)}_{k} \rmv,b_{k}^{(m)};\V{z}^{(m)}_{k} \big)$,
$\Psi^{j,m} \rmv\triangleq \Psi_{j,m}\big(a^{(j)}_{k} \rmv,b^{(m)}_{k}\big)$.}
  \label{fig:factorGraph}
\vspace{-5mm}
\end{figure}

\subsection{JPDA for Marginal MMSE-Estimation}
\label{sec:JPDA-Tracking}

JPDA methods are based on the MMSE estimation criterion. (An exception is the set JPDA (SJPDA) filter \cite{SveSve11}, which aims to minimize the mean optimal sub-pattern assignment metric.) In particular, the JPDA aims to calculate MMSE estimates for each individual target state $\RV{x}_k^{(j)}$, i.e.,
\begin{align}
\hat{\V{x}}_k^{(j)\text{MMSE}} = \ist\int \V{x}^{(j)}_k \ist f(\V{x}^{(j)}_k |\V{z}_{1:k})\ist\ist \text{d}\V{x}_k, \quad j = 1,\dots,l_k  \ist. \label{eq:JPDA_MMSE}
\end{align}
Contrary to MHT, in conventional JPDA it is assumed that the true sequence of existence variables $\hat{\V{r}}_{0:k}$ (and thus the number of targets $l_k$ at time $k$) is known\vspace{0mm}, i.e.,
\begin{equation}
f(\V{x}_{0:k},\V{a}_{1:k}|\V{z}_{1:k}) \propto f(\V{x}_{0:k}, \hat{\V{r}}_{0:k}, \V{a}_{1:k}|\V{z}_{1:k}). \label{eq:JPDA_Joint}
\vspace{0mm}
\end{equation}

The marginals needed for optimum state estimation in \eqref{eq:JPDA_MMSE} are then calculated from \eqref{eq:JPDA_Joint} according\vspace{1mm} to
\begin{align}
&f(\V{x}^{(j)}_k|\V{z}_{1:k}) = \ist\int \sum_{\V{a}_{1:k}} \ist f(\V{x}_{0:k},\V{a}_{1:k}|\V{z}_{1:k})\ist \text{d}\V{x}^{\sim(j)}_{0:k}\ist, \label{eq:JPDA_1}
\end{align}
where $\V{x}^{\sim(j)}_{0:k}\rmv$ denotes the vector that consists of all entries in $\V{x}_{0:k}\rmv$ except $\V{x}^{(j)}_{k}\rmv$.
 In practical implementations a track initialization and termination heuristic provides an approximate sequence of existence variables $\tilde{\V{r}}_{0:k}$. Under the assumption of linear and Gaussian measurement and motion models as used in the original formulation of the JPDA, approximate closed-form expressions for the marginalization \eqref{eq:JPDA_1} can be found. However, the summation over all joint associations hypotheses performed in \eqref{eq:JPDA_1} remains computationally infeasible in tracking scenarios with more than 7 targets. Further complexity reduction can be obtained by applying heuristic approximations \cite{Fit90,MusLaS08}, by using efficient hypothesis management \cite{horridge06}, or by limiting the number of target-measurement associations \cite{PatPop00}. 

A deficiency of JPDA tracking is known as track coalescence, which can arise when targets come close to each other. In such situations, JPDA is unable to maintain awareness of target positions and the estimated tracks tend to merge, and can become indistinguishable. Track coalescence is a direct consequence of calculating marginal distributions $f(\V{x}^{(j)}_k |\V{z}_{1:k})$ for the individual target states according to \eqref{eq:JPDA_1}. In particular, performing soft data association by means of the sum over all possible associations $\sum_{\V{a}_k}$ for each time step $k$, has the effect that targets with identical prior distribution will also have identical posterior distribution. Consequently, as targets are in close proximity and their prior distributions become similar, their posterior distributions tend to merge. Mitigation methods include the pruning of association hypotheses \cite{bloblo00,bloblo15}, approximate calculation of marginal pdfs via information-theoretic based optimization techniques \cite{SveSve11,Wil14} or mean-field approximations \cite{LauWil16}.

\subsection{BP for Marginal Detection and MMSE-Estimation}
\label{sec:BP-Tracking}
 
In BP-based multitarget tracking \cite{mey17,mey18,SolMeyBraHla:J19,MeyWil:J21}, we aim to compute the marginal posterior pdfs $f\big(\V{x}^{(j)}_{k},\V{r}^{(j)}_{k}\big|\V{z}_{1:k}\big)$ for each PT $j \in \{1,\dots,j_k\}$. From these marginal posterior pdfs, detection and MMSE estimation can be performed individually for each PT $j \in \{1,\dots,j_k\}$. 
 
  In particular, for each PT, marginal posterior probability mass functions $p\big(r^{(j)}_{k} \big| \V{z}_{1:k} \big)$ 
of the existence indicators $\rv{r}^{(j)}_{k}\rmv$ are
obtained from marginal posterior pdfs $f\big(\V{x}^{(j)}_{k},\V{r}^{(j)}_{k}\big|\V{z}_{1:k}\big)$
as
\begin{equation}
p\big(r^{(j)}_{k} \big| \V{z}_{1:k} \big) = \int \rmv f\big(\V{x}^{(j)}_{k}\rmv,r^{(j)}_{k} \big| \V{z}_{1:k}  \big) \ist \mathrm{d} \V{x}^{(j)}_{k}. \label{eq:detection}
\end{equation}
Targets are then declared to exist if $p\big(r^{(j)}_{k} \!\rmv=\!1\big|\V{z}_{1:k})$ is greater than a predefined threshold $P_{\text{th}}$. 
For all targets declared to exist, state estimation is performed by the MMSE estimator according\vspace{-1.5mm} to
\begin{equation}
\label{eq:BP-MMSE}
\hat{\V{x}}^{(j)\text{MMSE}}_{k} \ist \triangleq \ist\int \rmv \V{x}_k \ist \frac{f\big(\V{x}^{(j)}_{k}\rmv,r^{(j)}_{k} \!\rmv=\! 1 \big| \V{z}_{1:k} \big)}{p\big(r^{(j)}_{k} \!\rmv=\! 1 \big| \V{z}_{1:k} \big)} \ist \mathrm{d}\V{x}_k.
\vspace{0mm}
\end{equation}

Let us first investigate the calculation of the joint marginal pdfs $f\big(\V{x}_{k},r_{k}$ $\big| \V{z}_{1:k}\big)$ that involves the joint state and existence variable of all targets at time $k$. A naive marginalization of \eqref{eq:jointPosteriorComplete} can be performed\vspace{-.8mm} as
\begin{align}
&\hspace{0mm} f\big(\V{x}_{k},\V{r}_{k}\big| \V{z}_{1:k}\big) \nn \\[1.5mm]
&\hspace{0mm} = \int \sum_{\V{r}_{0:k-1}} \sum_{\V{a}_{1:k}} \ist\ist \sum_{\V{b}_{1:k}} \ist\ist f\big( \V{x}_{0:k}, \V{r}_{0:k}, \V{a}_{1:k}, \V{b}_{1:k} \big| \V{z}_{1:k} \big) \ist \mathrm{d} \V{x}_{0:k-1}. \nn \\[-3mm]
 \label{eq:marginalization1} \\[-7mm]
\nn
\end{align}

By using \eqref{eq:jointPosteriorComplete} in \eqref{eq:marginalization1} and marginalizing out the vectors $\V{x}_{1:k-2}$ $\V{r}_{1:k-2}$, $\V{a}_{1:k-1}$, and $\V{b}_{1:k-1}$, we obtain
\begin{align}
&f\big(\V{x}_k,r_{k}\big| \V{z}_{1:k}\big)  \nn\\[1mm]
&\hspace{6mm} = \int \sum_{\V{r}_{k-1}} \sum_{\V{a}_{k}} \ist\ist \sum_{\V{b}_{k}} \ist  f\big( \V{x}_{k-1}, \V{r}_{k-1} \big| \V{z}_{1:k-1} \big)   \nn\\[1mm]
&\hspace{12mm} \times g(\V{x}_{k}, \V{r}_{k}, \V{x}_{k-1}, \V{r}_{k-1}, \V{a}_{k}, \V{b}_{k}) \ist\ist \mathrm{d} \V{x}_{k-1}
 \label{eq:marginalization2} 
\end{align}
where $g(\V{x}_{k}, \V{r}_{k}, \V{x}_{k-1}, \V{r}_{k-1}, \V{a}_{k}, \V{b}_{k})$ is equal to the last three lines in \eqref{eq:jointPosteriorComplete} (with $k'$ replaced by $k$). Now marginals for individual targets used in \eqref{eq:detection} and \eqref{eq:BP-MMSE} can be obtained according to $f\big(\V{x}^{(j)}_k,r^{(j)}_{k}\big| \V{z}_{1:k}\big) = \int \sum_{ \V{r}^{\sim(j)}_{k}} f\big(\V{x}_k,r_{k}\big| \V{z}_{1:k}\big) \mathrm{d} \V{x}^{\sim(j)}_{k} \rmv\rmv$, \vspace{-.5mm} where $\V{r}^{\sim (j)}_{k}\rmv$ denotes the vector that consists of all\vspace{.3mm}  entries in $\V{r}_{k}\rmv$ except $\V{r}^{(j)}_{k}\rmv$ and $\V{x}^{\sim (j)}_{k}\rmv$ denotes the vector that consists of all entries in $\V{x}_{k}\rmv$ except $\V{x}^{(j)}_{k}\rmv$.

There are two interesting observations that can be made. (i) As in conventional filtering methods, by exploiting the temporal factorization structure of \eqref{eq:jointPosteriorComplete}, the computational complexity can be strongly reduced. (Note that, e.g., the summation over $\sum_{\V{r}_{0:k-1}}$ in  \eqref{eq:marginalization1} consists of $2^{\hspace{.1mm} \prod^{k-1}_{k'=0} j_{k'} }$ terms, while\vspace{-.1mm} the $k$ individual summations performed sequentially according to  \eqref{eq:marginalization2} only consist of $\sum^{k-1}_{k'=0} \ist 2^{\hspace{.1mm} j_{k'}}$ terms.) (ii) The optimal MMSE solution is a ``single scan solution'' in the sense that the optimal marginal \ac{pdf} $f\big(\V{x}_k,r_{k}\big| \V{z}_{1:k}\big)$ at time $k$ can be directly obtained from the optimal marginal \ac{pdf} $f\big(\V{x}_{k-1},r_{k-1}\big| \V{z}_{1:k-1}\big)$ at time $k-1$, i.e., $f\big(\V{x}_{k-1},r_{k-1}\big| \V{z}_{1:k-1}\big)$ provides all relevant information from past time steps.

BP makes it possible to also exploit the spatial factorization structure of \eqref{eq:jointPosteriorComplete} to further reduce computational complexity \cite{mey17,mey18}. It operates on a factor graph that represents the statistical model of the considered estimation problem \cite{kschischang01}. In particular, marginal pdfs are calculated efficiently by performing local operations for the graph nodes and exchanging their results called ``messages'' along the graph edges. Since the factor graph in Fig.~\ref{fig:factorGraph} representing this structure has cycles, BP can only provide approximate marginal pdfs  $f\big(\V{x}^{(j)}_{k},\V{r}^{(j)}_{k}\big|\V{z}_{1:k}\big)$, $j \in \{1,\dots,j_k\}$ \cite{kschischang01}. These approximate marginal pdfs are typically referred to as ``beliefs''.  Furthermore, the cycles in the factor graph lead to overconfident messages and beliefs \cite{WeiFre:01}, i.e.,  their spread ``downplays'' their true uncertainty.

Since BP performs marginalization similarly to JPDA, it is expected to suffer from track coalescence. However, as will be seen in Section \ref{sec:results}, track coalescence is reduced significantly compared to JPDA. This can be explained by the particle-based implementation used by BP-based MTT that makes it possible to more accurately represent multimodal marginal \acp{pdf}. Furthermore, the overconfident nature of BP-based data association ``upsells'' the most likely measurement-to-target associations hypothesis and thus leads to a hybrid form of soft and hard data association and further reduces \vspace{-.85mm} coalescence.

 \begin{figure*}[t!]
\centering
\hspace{3mm}
\subfloat[\label{fig:True1}]{\scalebox{0.3}{\hspace{-12mm} \includegraphics[scale=.9]{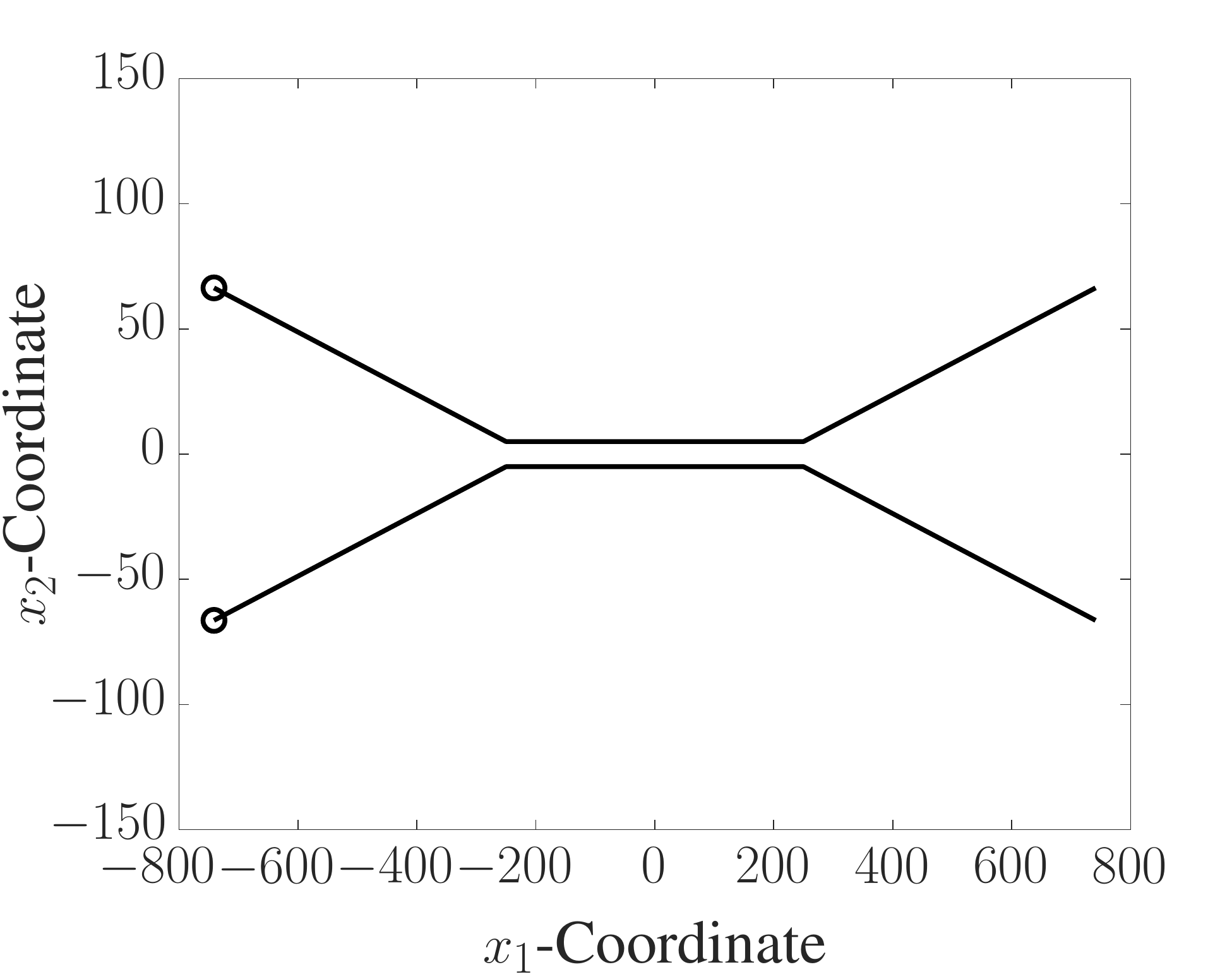}}}
\hspace{10mm} 
\subfloat[\label{fig:True2}]{\scalebox{0.3}{\hspace{-12mm} \includegraphics[scale=.9]{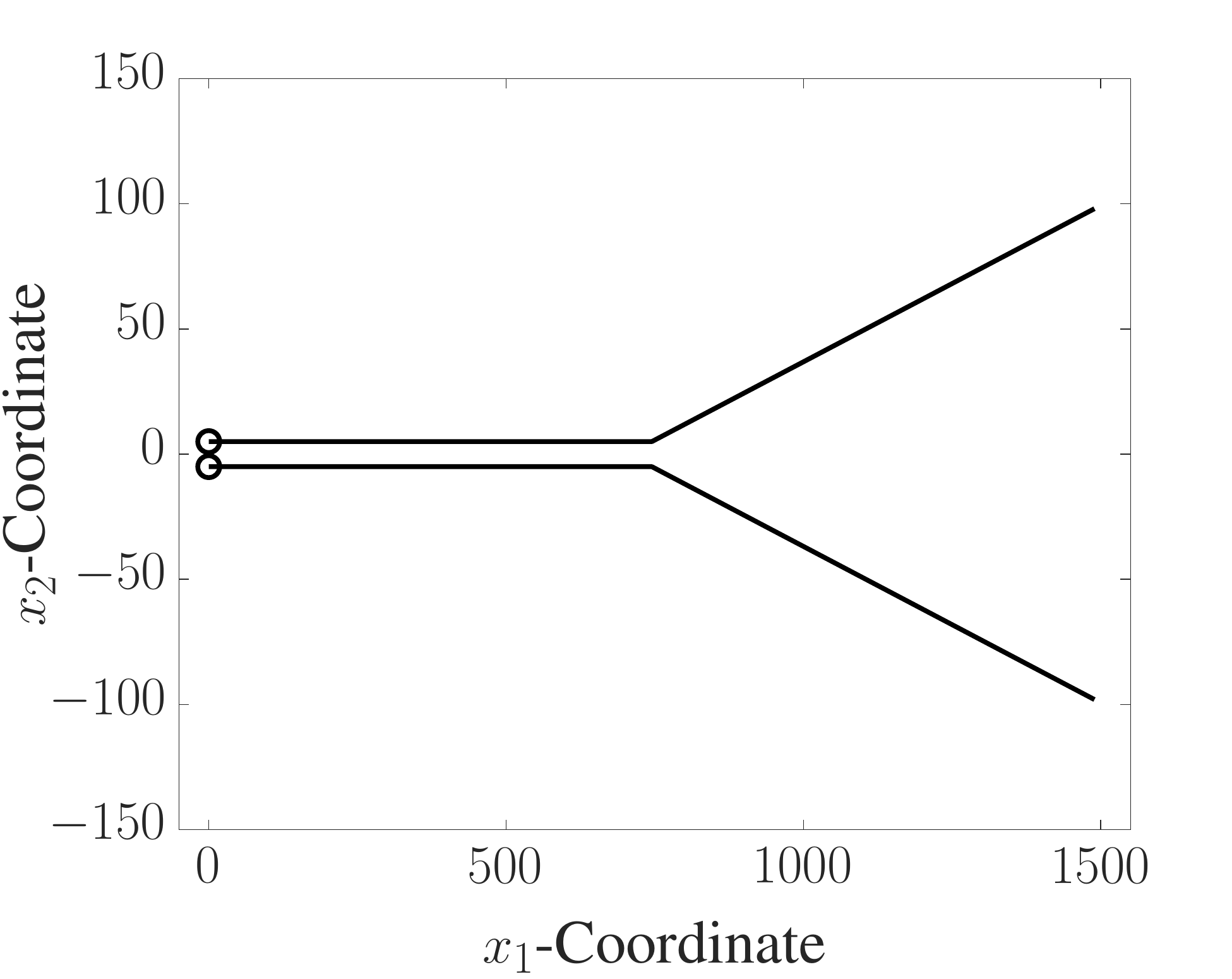}}}
\hspace{10mm}        
\subfloat[\label{fig:True3}]{\scalebox{0.3}{\hspace{-12mm} \includegraphics[scale=.9]{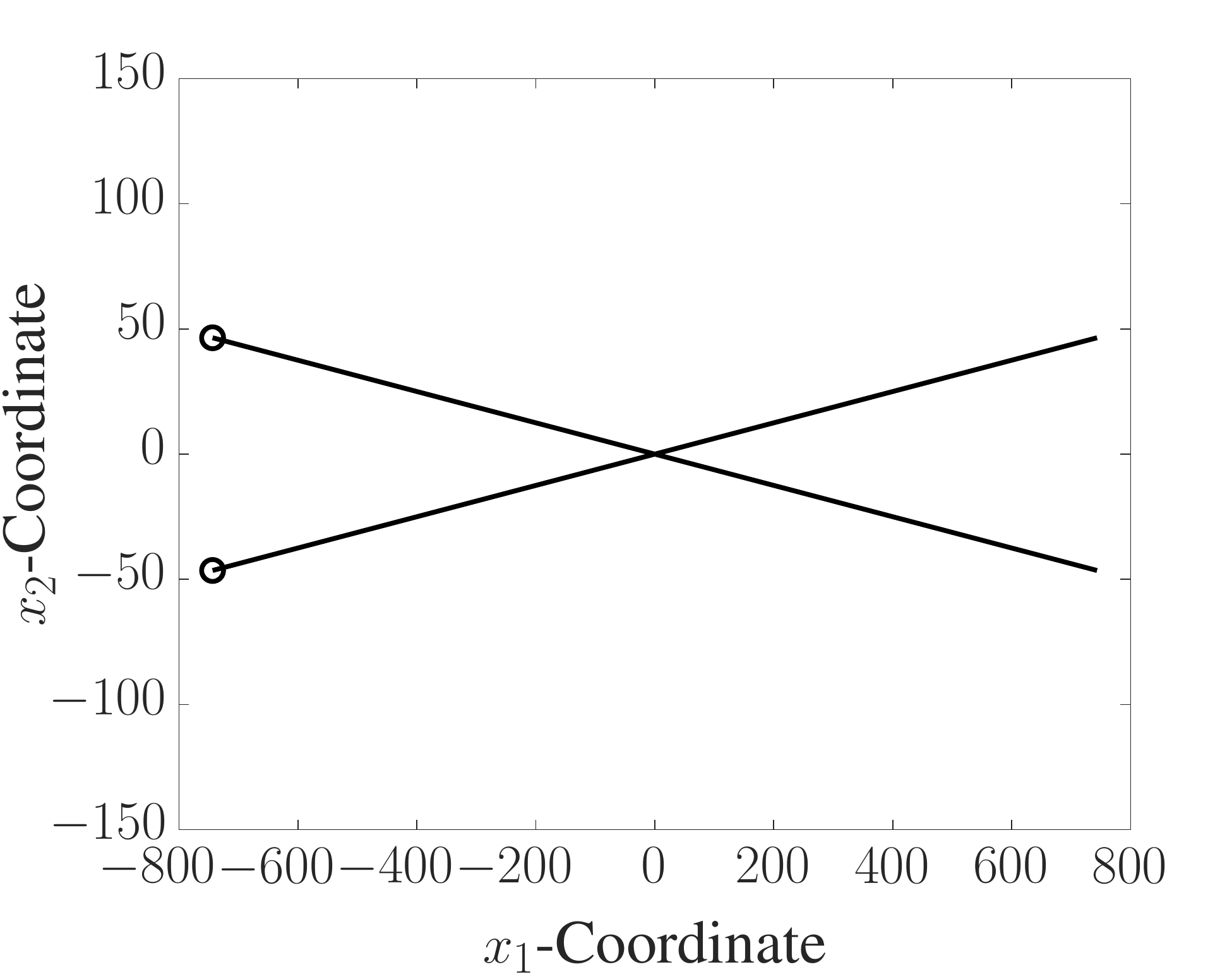}}}
\captionsetup{singlelinecheck = false, justification=justified}
\vspace{-2mm}
\caption{True target tracks used in scenario 1 (a), scenario 2 (b), and scenario 3 (c). Initial target locations are indicated by a black circle. Note that the two coordinate axis are scaled differently.}
\label{fig:scenarios}
\vspace{-8mm}
\end{figure*}

 \begin{figure}[t!]
\centering
\vspace{2mm}
\subfloat[\label{fig:ospaError}]{\scalebox{0.3}{\hspace{-16mm} \includegraphics[scale=.9]{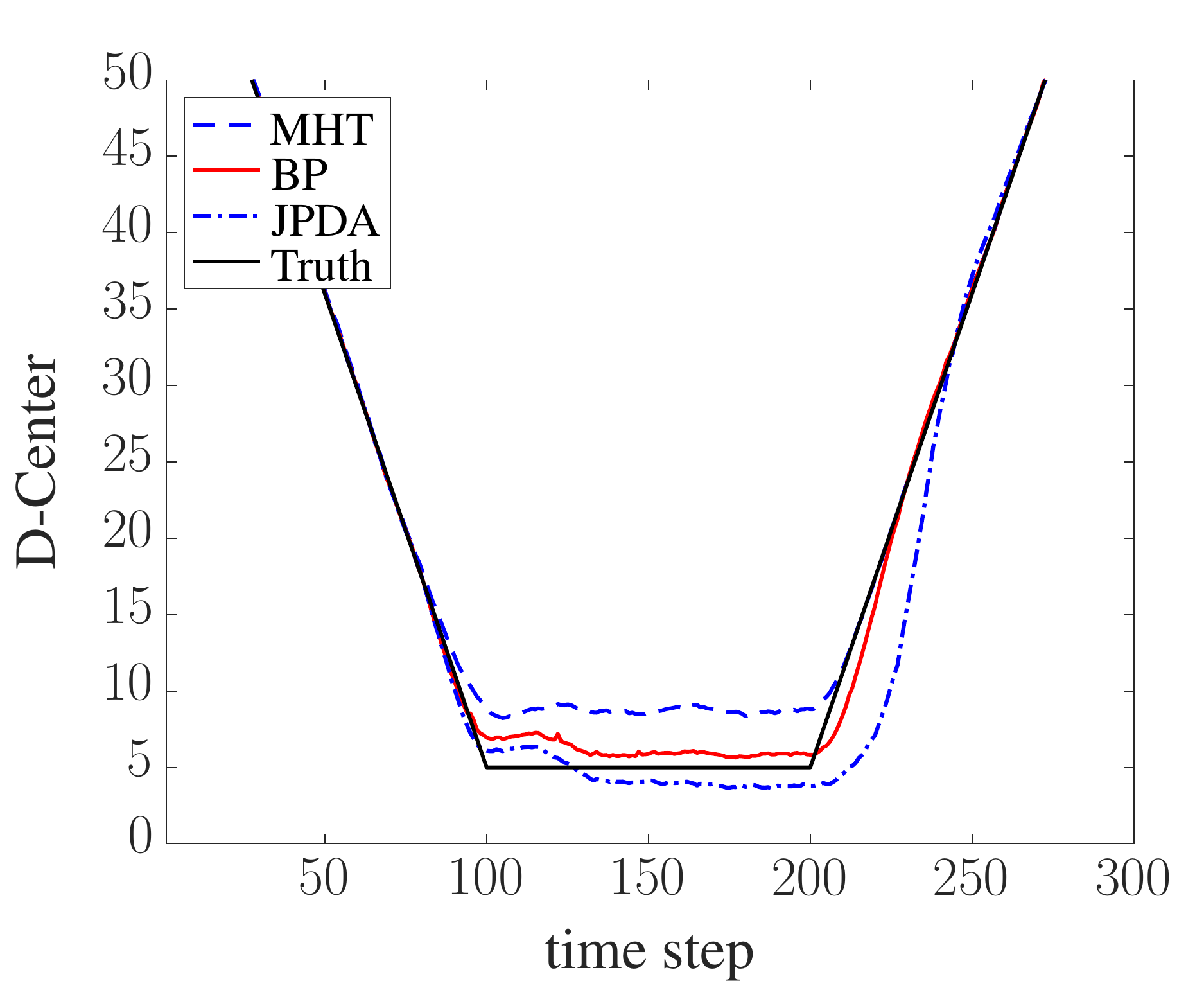}}}
\vspace{0mm}
\subfloat[\label{fig:cardinalityError}]{\scalebox{0.3}{\hspace{-16mm} \includegraphics[scale=.9]{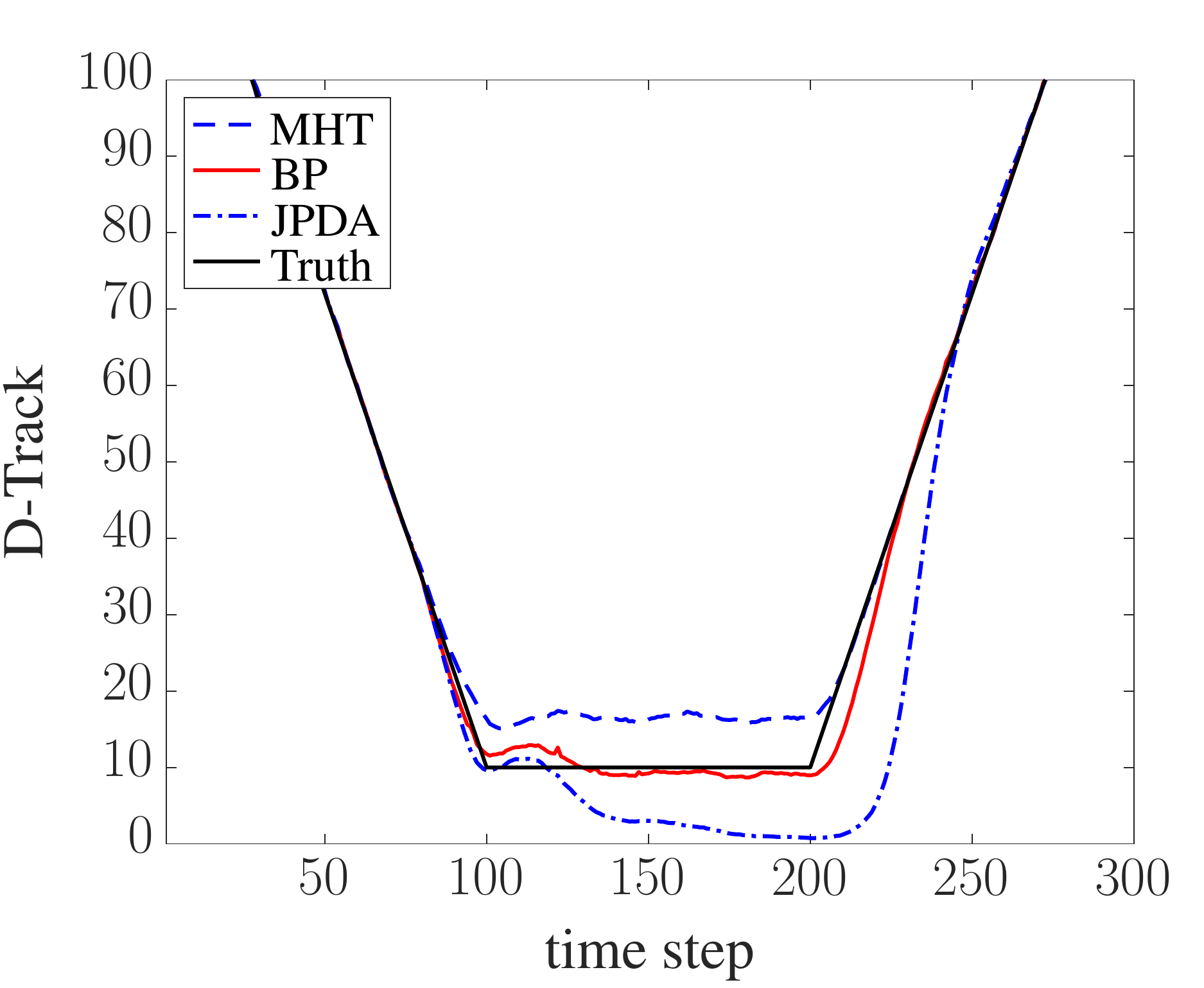}}}
\captionsetup{singlelinecheck = false, justification=justified}
\caption{Scenario 1: D-Center (a) and D-Tracks (b) for the three considered multitarget tracking methods.}
\label{fig:distanceTracks}
\vspace{-6mm}
\end{figure}

\section{Results}\label{sec:results}
\vspace{-.5mm}

We simulate three different scenarios with two targets whose states consist of two-dimensional (2D) position and velocity, i.e., 
$\RV{x}_{k}^{(j)} \rmv= [\rv{x}_{1,k}^{(j)} \;\ist \rv{x}_{2,k}^{(j)} \;\ist \dot{\rv{x}}_{1,k}^{(j)} \;\ist \dot{\rv{x}}_{2,k}^{(j)}]^{\text{T}}\rmv$, $j\rmv\in\rmv\{1,2\}$. The targets are assumed to move according to the near constant-velocity motion model, i.e., $\RV{x}_{k}^{(j)} = \M{A}\ist\RV{x}_{k-1}^{(j)} + \RV{u}_{k}^{(j)}$ where $\RV{u}_{k}^{(j)} \rmv\sim \Set{N}(\V{0},\M{\Sigma}_{\V{u}} )$  is a sequence of independent and identically distributed (iid) 2D Gaussian random vectors. The matrixes $\M{A}$ and $\M{\Sigma}_{\V{u}}$ represent discretized continuous-time kinematics \cite{barShalom02}, i.e.,
\begin{equation}
\label{eq:sysmat}
\small
\M{A} = 
{ \begin{pmatrix}
  1 & 0 & T & 0 \\
   0 & 1 & 0 & T \\
   0  & 0  & 1 & 0 \\
   0 & 0 & 0 & 1
  \end{pmatrix} } \ist,
\quad\, \M{\Sigma}_{\V{u}} =
 { \begin{pmatrix}
   \frac{T^3}{3} & 0 & \frac{T^2}{2} & 0\\
   0 & \frac{T^3}{3} & 0 & \frac{T^2}{2}\\
   \frac{T^2}{2} & 0 & T & 0\\
   0 & \frac{T^2}{2} & 0 & T
  \end{pmatrix} } \rmv \sigma^2_{\V{u}}\nn
  \vspace{-1mm}
\end{equation}
with $\sigma^2_{\V{u}} \!=\rmv 0.1 \ist\ist \text{m}^2/\text{s}^4$ and $T\rmv=\rmv1\ist\text{s}$. The true target tracks for the three different scenarios as well as the initial target positions are shown in Fig.~\ref{fig:scenarios}. Note that in scenarios 1 and 2, the minimum distance between the two targets is equal to $\sigma_{\V{v}}$ (defined in what follows). The region-of-interest (ROI) is given by $[-750\ist\text{m}, \ist 750\ist\text{m} ]  \times [-750\ist\text{m}, \ist 750\ist\text{m}]$ for scenarios 1 and 3 and by $[0\ist\text{m}, \ist 1500\ist\text{m} ]  \times [-750\ist\text{m}, \ist 750\ist\text{m}]$ for scenario 2.

A single sensor performs noisy measurements of the positions of the targets. More specifically, the target-generated measurements are given\vspace{-.5mm} by
\[
\RV{z}^{(m)}_{k} =\ist   [\rv{x}_{1,k}^{(m)} \;\ist \rv{x}_{2,k}^{(m)} ]^{\text{T}} \!+\ist \RV{v}_{k}^{(m)}
\vspace{-.8mm}
\]
where $\RV{v}_{k}^{(m)} \sim \Set{N}(\V{0},\sigma_{\V{v}}^2 \M{I}_2)$ with $\sigma_{\V{v}} \rmv=\rmv 10$m is an iid sequence of 2D Gaussian random vectors. 
The clutter pdf $f_{\mathrm{c}}\big( \V{z}_{k}^{(m)} \big)$ is uniform on the ROI. The mean number of clutter measurements is $\mu_{\mathrm{c}} \!\rmv=\! 10$. A birth distribution $f_{\mathrm{b}}(\V{x}_{n,k})$ that is uniform on the ROI is considered. Furthermore, the mean number of newborn targets per time step and the survival probability are set to $\mu_{\mathrm{b}} \rmv=\rmv 0.01$ and $p_{\mathrm{s}} = 0.995$, respectively. 

We use the particle-based implementation of BP for MTT presented in \cite{mey17} with $J \rmv=\rmv 5000$ particles for each PT state. The threshold for target confirmation is $P_{\text{th}} \rmv=\rmv 0.5$. Potential target states are pruned if their existence probability is below threshold $P_{\text{pr}} = 10^{-5}$. We simulated 300 time steps $n$ and 1000 Monte Carlo realizations for each of the three scenarios. 
We used the track-oriented variant of the MHT method. The JPDA and the MHT methods used a gate validation threshold  of 13.82, which corresponds to a probability of 0.999 that target-originated measurements (when the corresponding target exists) 
are within the gate. A gate validation threshold of 9.21, which corresponds to a probability of 0.99 that target-originated measurements are within the gate, led to similar results (not shown). The track confirmation logics of JPDA and MHT are set to 10/16 and 12/24, respectively. Furthermore, the JPDA and the MHT allow at most 13 missed detections prior to track termination. These values are tuned to best represent the algorithms across all three scenarios.
The hypothesis depth of the MHT filter is set to five scans. 

To  visualize the coalescence and repulsion effects related to the estimated tracks provided by the JPDA, MHT, and BP methods, we compute the mean distance of the estimated y-coordinates to the center between the two tracks' (``D-Center'') as well as the distance between the estimated y-coordinates of the two tracks (``D-Tracks''). In particular the D-Center is the mean of $(|\hat{x}_{2,k}^{(1)}| + |\hat{x}_{2,k}^{(2)}|)/2$, averaged over all time steps and realizations. If more than two tracks are estimated at a certain time step, then only the two estimates which are closest (in a global nearest neighbor sense) to the true tracks are considered. Similarly, if one or no tracks are estimated at certain time steps, the missing state estimates are excluded from the mean. Furthermore, D-Tracks is the mean of $|\hat{x}_{2,k}^{(1)} - \hat{x}_{2,k}^{(2)}|$, averaged over all time steps and iterations. Additional and missing tracks are handled similarly as for D-Center. D-Tracks can alert to coalescence, while (assuming tracks symmetric about the origin) D-Center examines the quality of the tracks' centroid. For example, if both are zero, the tracks may have coalesced but at least their joint track is good; but if D-Center is large then even the joint track is poor.

Fig.~\ref{fig:distanceTracks} shows D-Center and D-Tracks for scenario 1. In Fig.~\ref{fig:distanceTracks}(a) it can be seen that when the two targets move in parallel and in close proximity, the tracks provided by the MHT yield an increased D-Center compared to the true tracks due to track repulsion. On the other hand, for tracks provided by the JPDA, D-Center is reduced; for time steps $200 \leq n \leq 250$, the reduced D-Center due to track coalescence is particularly pronounced after the true tracks have split again. Furthermore, it can also be observed that the tracks provided by BP only suffer from a very weak form of repulsion (when true tracks are in close proximity) and coalescence (after the true tracks have split). For the D-Tracks in \ref{fig:distanceTracks}(b), it can be seen that the track coalescence effect leads to target tracks that move towards each other until they have almost identical y-coordinates. In scenarios 2 and 3, we have observed a similar behavior of D-Center and D-Tracks (not shown).
 
\begin{figure}[t!]
\centering
\includegraphics[scale=.28]{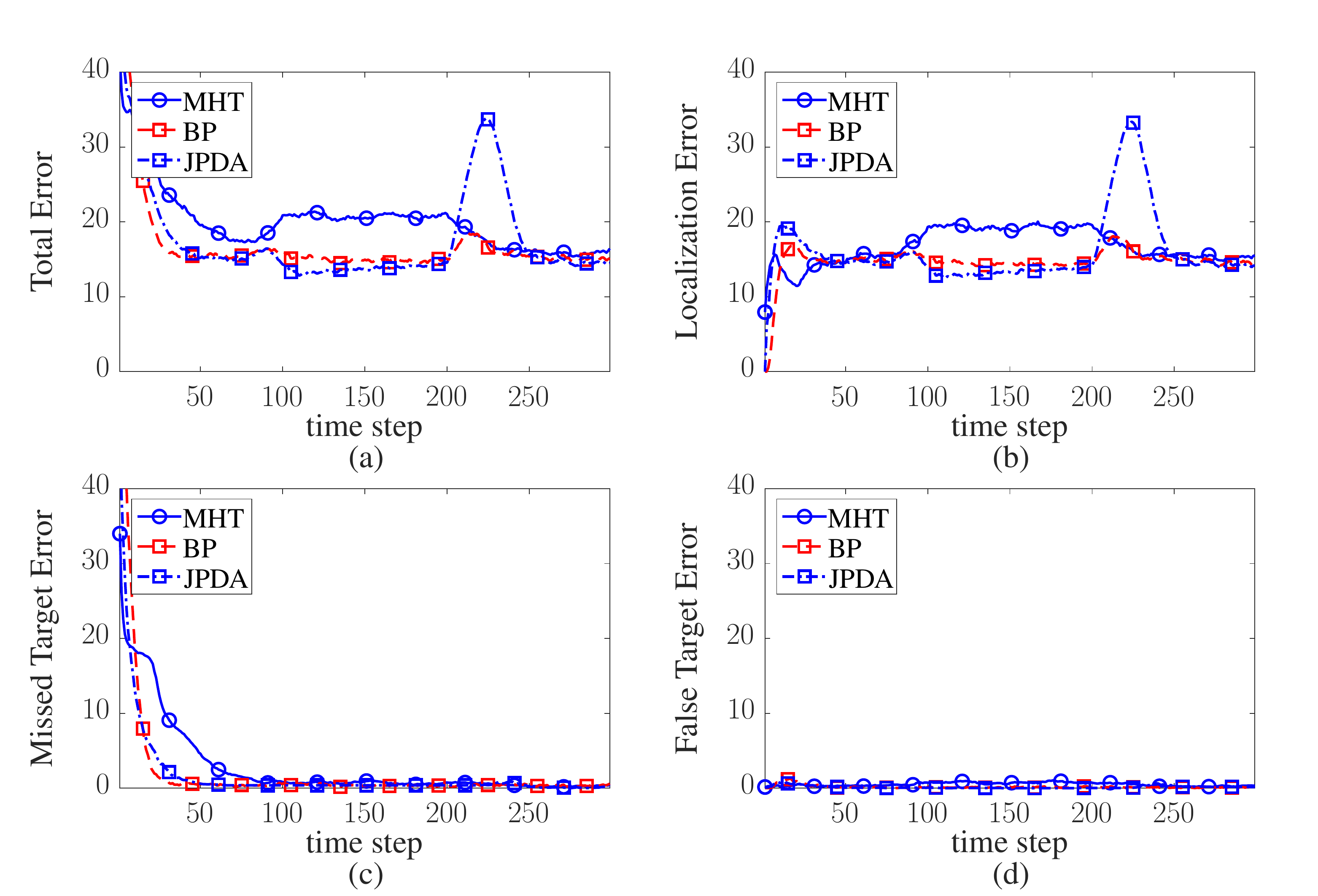}
\caption{Scenario 1: Mean total \ac{gospa} (a) mean localization error (b), mean missed target error (c), and mean false target error (d) versus time for the three considered multitarget tracking methods.}
\label{fig:scenario1GOSPA}
\vspace{-5mm}
\end{figure}

As an additional performance metric, we use the \ac{gospa} \cite{RahGarSve:C17} based on the 2-norm as inner metric as well as parameters $p\rmv=\rmv1$ and $c\rmv=\rmv50$. The mean \ac{gospa}---averaged over all simulation runs---for the three tracking methods in scenario 1 is shown in Fig.~\ref{fig:scenario1GOSPA}. We show the mean total \ac{gospa} in Fig.~\ref{fig:scenario1GOSPA}(a) as well as the individual \ac{gospa} contributions, namely the mean total localization error, the mean missed target error, and the mean false target error in Figs.~\ref{fig:scenario1GOSPA}(b)--(d), respectively. The estimated MHT tracks yield a slightly increased total \ac{gospa} for times $n\rmv< 100$, due to an increase number of missed tracks compared to the estimated JPDA and BP tracks. This slightly increased number of missed tracks is the result of MHT parameters that were tuned to yield satisfying performance in all three scenarios. (Note that for the JPDA method it was not possible to find parameters that lead to a satisfying performance in scenario 2.) For the times where the targets are in close proximity, i.e., $100 \leq n \leq 200$, coalescence of estimated JPDA tracks results in a lower localization error compared to the other time steps and compared to the other methods. However, right after the true target tracks split, i.e., at times $200 < n < 240$, the JPDA yields a significant localization error peak. Due to track repulsion, the localization error of estimated MHT tracks is increased when targets are in close proximity. Notably, the localization error related to BP is significantly reduced compared to the other two methods.

\begin{figure}[t!]
\centering
\includegraphics[scale=.28]{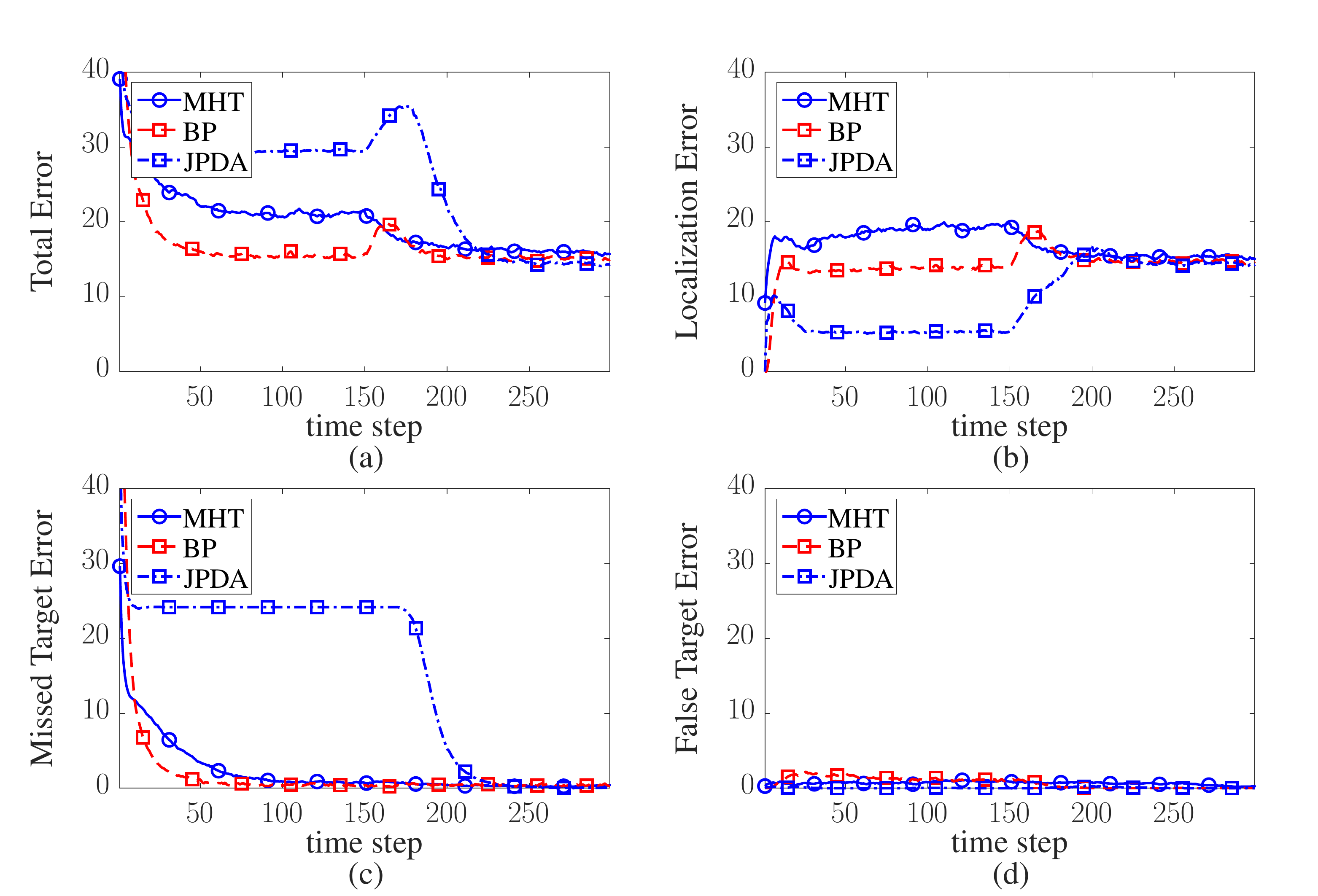}
\caption{Scenario 2: Mean total \ac{gospa} (a) mean localization error (b), mean missed target error (c), and mean false target error (d) versus time for the three considered multitarget tracking methods.}
\label{fig:scenario2GOSPA}
\vspace{-6mm}
\end{figure}

Fig.~\ref{fig:scenario2GOSPA} shows the mean \ac{gospa} error of the JPDA, MHT, and BP methods in scenario 2. For time steps where targets are in close proximity, i.e., $0 \leq n \leq 150$, due to track repulsion, the localization and total \ac{gospa} errors of the MHT are increased compared to those of the BP method. In addition, it can be seen that for times $n < 150$, the missed target and total \ac{gospa} errors of the JPDA result are significantly increased compared to the other two methods. This is because the confirmation logic of the JPDA is unable to initialize two tracks as long the true target states remain in close proximity \cite{barShalom11}. The fact that the JPDA only creates a single track for these times steps, results in a reduced localization error compared to MHT and BP. As expected this localization error is 5m on average, which is equal to half the distance between the two tracks. In scenario 2, BP yields a slightly increased false tracks error compared to JPDA and MHT. This increased error is mostly due the generation of short tracks that only consists of few time steps and could be removed in post-processing. (Note that the tracking confirmation logics used for JPDA and MHT automatically suppress such short tracks.)

 \begin{figure}[t!]
\centering
\includegraphics[scale=.28]{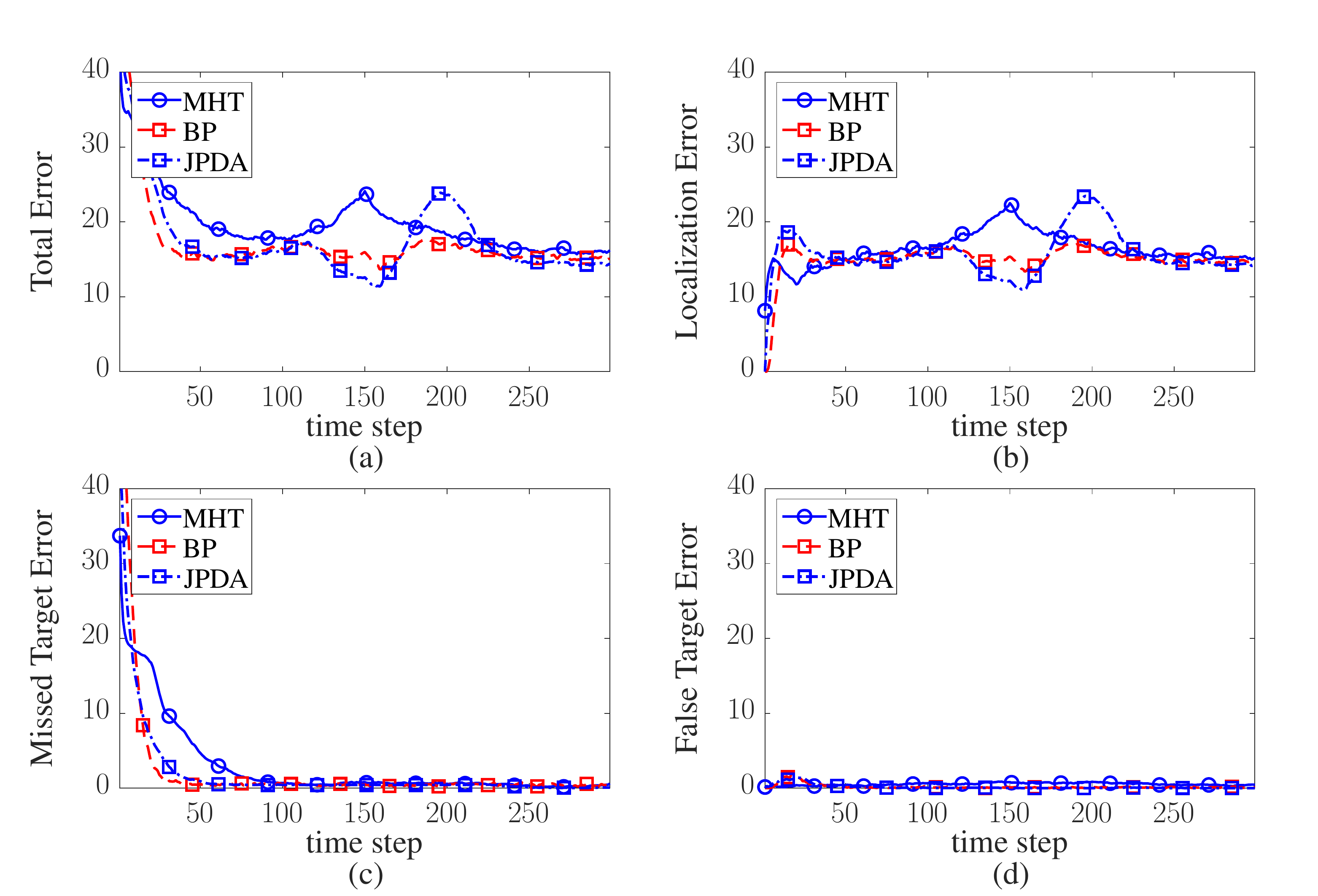}
\caption{Scenario 3: Mean total \ac{gospa} (a) mean localization error (b), mean missed target error (c), and mean false target error (d) versus time for the three considered multitarget tracking methods.}
\label{fig:scenario3GOSPA}
\vspace{-5mm}
\end{figure}

Finally, Fig.~\ref{fig:scenario3GOSPA} shows the mean \ac{gospa} error of the JPDA, MHT, and BP methods in scenario 3. Here we can again see that for time steps where targets are in close proximity, i.e., $ 125 \leq n \leq 175$, due to track repulsion, the localization and total \ac{gospa} error of MHT is increased compared to the other methods. On the other hand, track coalescence of the JPDA decreases the localization and total \ac{gospa} error when targets are in close proximity ($ 125 \leq n \leq 175$) and increases them right after the two targets have split ($ 175 \leq n \leq 225$). An important aspect in a scenario with crossing tracks are track swaps \cite{Willet07}. A quantification of track swaps based on track-level metrics \cite{CorGriDet:C06} will be reported in future\vspace{-2mm} work.

\section{Conclusion}\label{sec:conclusion}
\vspace{-1mm}

In this paper, we reviewed the estimation strategies of multiple hypothesis tracking (MHT),  joint probabilistic data association (JPDA), and the recently introduced belief propagation (BP) approach for MTT. Since BP performs marginalization similarly to JPDA, it is expected to suffer from track coalescence. Our results however indicated that BP-based MTT can mostly avoid coalescence and repulsion, problems that respectively afflict JPDA and MHT. This can be explained by the particle-based implementation used by BP-based MTT that makes it possible to accurately represent multimodal marginal \acp{pdf}. Furthermore, the overconfident nature of BP-based data association ``upsells'' the most likely measurement-to-target associations hypothesis and thus leads to a hybrid form of soft and hard data association that reduces track\vspace{0mm} coalescence.
\vspace{-2mm}

\renewcommand{\baselinestretch}{.946}
\footnotesize
\selectfont
\bibliographystyle{IEEEtran}
\bibliography{references,IEEEabrv,StringDefinitions}

\end{document}